%% file: main.tex
\documentclass{Interspeech2024}

\usepackage{subcaption}
\usepackage{multirow}
\usepackage[table]{xcolor}
\usepackage{array}
\usepackage{amsmath}
\usepackage{amssymb}

\interspeechcameraready

\title{An Exploration of Length Generalization in Transformer-Based Speech Enhancement}

\name[affiliation={1,2}]{Qiquan}{Zhang}
\name[affiliation={4,*}]{Hongxu}{Zhu}
\name[affiliation={5,*}]{Xinyuan}{Qian}
\name[affiliation={1}]{Eliathamby}{Ambikairajah}  
\name[affiliation={3,2}]{Haizhou}{Li}

\address{
  $^1$The University of New South Wales, Australia, 
  $^2$National University of Singapore, Singapore \\
  $^3$SDS, SRIBD, The Chinese University of Hong Kong, Shenzhen, China \\
  $^4$ ASTRI, Hong Kong, $^5$University of Science and Technology Beijing, China}
\email{\{zhangqiquan\_hit;e.ambikairajah\}@unsw.edu.au, hongxuzhu@astri.org}

\keywords{\textcolor{black}{speech enhancement, Transformer, length generalization, position embedding}}

\begin{document}

\maketitle

\begin{abstract}
\textcolor{black}{The use of Transformer architectures has facilitated remarkable progress in speech enhancement. Training Transformers using substantially long speech utterances is often infeasible as self-attention suffers from quadratic complexity. It is a critical and unexplored challenge for a Transformer-based speech enhancement model to learn from short speech utterances and generalize to longer ones. In this paper, we conduct comprehensive experiments to explore the length generalization problem in speech enhancement with Transformer. Our findings first establish that position embedding provides an effective instrument to alleviate the impact of utterance length on Transformer-based speech enhancement. Specifically, we explore four different position embedding schemes to enable length generalization. The results confirm the superiority of relative position embeddings (RPEs) over absolute PE (APEs) in length generalization.\footnote{\textcolor{black}{This work is supported by ARC Discovery Grant DP1900102479; NSFC Grant No.~62271432; Internal Project Fund from Shenzhen Research Institute of Big Data (Grant No.~T00120220002); Shenzhen Science and Technology Research Fund (Fundamental Research Key Project Grant No.~JCYJ20220818103001002. * Corresponding author}} 
}

\end{abstract}

\section{Introduction}
\textcolor{black}{Speech enhancement seeks to reconstruct the clean speech signal from the noisy mixture contaminated by surrounding background noise, with a wide range of applications including hearing aids, audio-video conference, and automatic speech recognition (ASR). Traditionally, many approaches leverage the assumed statistical properties of noise and speech signals to derive an estimator of clean speech~\cite{loizou2013,zhang2019,zhang2019fast}. These approaches fail to eliminate rapidly varying noise signals.}

\textcolor{black}{In recent years, the field of speech enhancement has been dominated by supervised learning methods that optimize deep neural networks (DNNs) to perform a non-linear mapping from the noisy speech to the designed objective~\cite{wang2018supervised,mamba}. These methods mainly fall into waveform domain~\cite{realse,cleanunet} and spectral domain schemes. Waveform domain methods often utilize a DNN with an explicit encoder-decoder architecture to directly separate the clean waveform given a noisy waveform input. Spectral domain methods, in contrast, first transform the waveform representation into spectral representation, such as magnitude spectrum (MS)~\cite{wang2018supervised}, complex spectrum (CS)~\cite{tan2019learning}, and log-power spectrum (LPS)~\cite{xu2014regression}. Then, DNNs are trained to estimate the spectral feature of clean speech, or a spectral mask such as the ideal ratio mask (IRM)~\cite{wang2014training}, spectral magnitude mask (SMM)~\cite{wang2014training}, complex ideal ratio mask (cIRM)~\cite{williamson2015complex}, and phase-sensitive mask (PSM)~\cite{erdogan2015phase}.}

\begin{figure}[!ht]
% \vskip 0.1in
\centering
\centerline{\includegraphics[width=0.89\columnwidth]{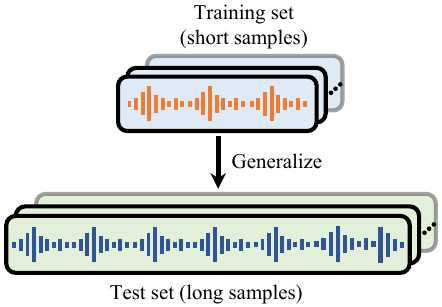}}
\caption{\textcolor{black}{Length generalization in speech enhancement: Train on short speech samples and inference on long speech samples. 
% The ability to learn from short speech samples (training set) to generalize to longer speech samples (test set).
} }
\label{fig1}
\vspace{-2.5em}
\end{figure}

Numerous network architectures have been employed for speech enhancement. Multi-layer perceptrons (MLPs) are only capable of leveraging local information with a context window. With the capability to model the long-range dynamic dependencies of speech signals, long short-term memory (LSTM) networks have been exploited for speech enhancement~\cite{chenlstm,weninger2015speech}. However, recurrent neural networks (RNNs) suffer from the sequential modeling nature, excluding their use in some applications. Subsequently, temporal convolution networks (TCNs) are successfully applied for speech enhancement and demonstrate impressive performance~\cite{zhang2021temporal,tfaj,TCNN,zhang2019monaural,TFA}. TCNs utilize stacked dilated 1-D convolution layers to build a large receptive field, which allows the model to leverage long-term contextual dependencies~\cite{TCN2018}. Transformer architectures have recently demonstrated state-of-the-art results across various signal processing tasks, including speech enhancement~\cite{ripple,sepformerstft,tgsa,10446337} and image recognition~\cite{ren2023sg}. As the core piece in Transformer, the multi-head self-attention (MHSA) module learns the representation of each element by capturing the interaction amongst all elements in a sequence. It grants Transformers the capability to effectively model global contextual information.

In supervised learning tasks, the ability of models to generalize to cases unseen at training time is critical. Extensive works have been conducted to study the generalization performance of neural speech enhancement models on unseen noise, speaker, and SNR conditions. Since self-attention has quadratic complexity, it is infeasible to train Transformers on all possible speech lengths. 
% \textcolor{blue}{Speech enhancement task often requires the model to continue to perform well on longer noisy inputs than ones encountered at the training time.} 
Hence, it is desirable to enable length generalization: ``Can a Transformer-based speech enhancement model generalize from short speech signals to longer ones?'' See Fig.~\ref{fig1} for an illustrative example of length generalization in speech enhancement. \textcolor{black}{Position embedding appears to be a primary factor in the length generalization of Transformers. There have been some attempts to explore position embedding to enable Transformer-based language models with the length generalization property~\cite{alibi,RoPE}, such as T5~\cite{T5} and KERPLE~\cite{chi2022kerple}. To our knowledge, the length generalization problem in Transformer speech enhancement has not yet been explored. This paper is the first to study the pathologies that occur when Transformers are asked to generalize to enhance longer noisy inputs. We explore four advanced position embedding methods to enable length generalization for Transformer speech enhancement and show that relative position embedding (RPE) is more robust to input length change than absolute position embedding (APE).} 

This paper is organized as follows. We formulate the problem of neural time-frequency speech enhancement in Section~\ref{sec:2}. We present positional encoding methods in Section~\ref{sec:3}. Section~\ref{sec:4} presents speech enhancement with position-aware Transformers. Section~\ref{sec:5} introduces the experimental setup and analyzes the results. Section~\ref{sec:6} concludes this paper.

\vspace{-0.5em}
\section{Problem Formulation}\label{sec:2}

\textcolor{black}{Given the clean speech waveform $\bm{s}\in\mathbb{R}^{1\times N}$ and additive noise waveform $\bm{d}\in\mathbb{R}^{1\times N}$, the observed noisy waveform $\bm{y}$ can be formulated as $\bm{y}=\bm{s}+\bm{d}$, with $N$ denoting the number of time samples. The short-time Fourier transform (STFT) is applied to transform the raw waveform into the time-frequency representation (or spectrogram), given by $Y_{t,k}=S_{t,k}+D_{t,k}$, where $Y_{t,k}$, $S_{t,k}$, and $D_{t,k}$ respectively represents complex-valued STFT spectra of the noisy mixture, clean speech, and noise, at the time frame $t$ and the frequency bin $k$. One typical neural speech enhancement scheme is to optimize a DNN model to predict a spectral mask $\widehat{M}_{t,k}$ from the noisy spectral magnitude. Then, the clean spectrum is obtained by applying the spectral mask to the noisy spectrum, given as $\widehat{S}_{t,k}=Y_{t,k}\cdot \widehat{M}_{t,k}$. In this work, we adopt the commonly used phase-sensitive mask (PSM) to carry out speech enhancement experiments for the exploration of length generalization. The PSM is formulated as~\cite{erdogan2015phase}:
\begin{equation}\label{PSM}
\setlength{\abovedisplayskip}{5pt}
\setlength{\belowdisplayskip}{5pt}
\text{PSM}_{t,k}=\frac{\left|S_{t, k}\right|}{|Y_{t, k}|}\cos\left(\Phi_{S_{t,k}-{Y_{t, k}}}\right)
% \vspace{-0.1em}
\end{equation}
where $|\cdot|$ extracts the magnitude spectrum, and $\Phi_{S_{t,k}-Y_{t, k}}$ denotes the difference between the clean and noisy spectral phase.}

\section{Position Embedding}\label{sec:3}
% \vspace{-1.0em}
% \subsection{Overall Architecture}
\textcolor{black}{There is a variety of position embedding methods explored to encode position information into Transformers, which can fall into absolute position embedding (APE) and relative position embedding (RPE).}
\vspace{-0.5em}
\subsection{Absolute Position Embedding}
\vspace{-0.5em}
\textcolor{black}{APE assigns a unique position embedding $\textbf{P}_{t}$ to each position $t$ in a sequence. The input embeddings are summed with $\textbf{P}$ to form the input fed to the actual Transformers. The two common variants of APE are fixed or learnable position embedding.}

\textbf{Sinusoidal Position Embedding}.~\textcolor{black}{In the original Transformer~\cite{transformer}, the fixed sinusoidal functions of different frequencies are used to compute the position embeddings. The position embedding $\textbf{P}_{t}$ assigned to the position $t$ is defined as 
\begin{equation}
\setlength{\abovedisplayskip}{3pt}
\setlength{\belowdisplayskip}{3pt}
\textbf{P}_{t,d}\!=
\begin{cases}
\sin\left(10000^{-\frac{d}{d_{model}}}\cdot t\right), & \text{if } d \text{ is even } \\
\cos\left(10000^{-\frac{(d-1)}{d_{model}}} \cdot t\right), & \text{if } d \text{ is odd }
\end{cases}
\end{equation}
where $d_{model}$ denotes the dimension of input embeddings, with $d=\left\{1,...,d_{model}\right\}$ and $t=\left\{1,..,T\right\}$.}

\textbf{Learnable APE}. \textcolor{black}{Alternatively, the position embeddings can be learnable~\cite{kenton2019bert}, where each position is assigned a fully learnable position vector. They are randomly initialized and jointly updated with the model's parameters.}
\vspace{-0.8em}

% ~\textcolor{blue}{In The learnable position embeddings are used in models such as BERT~\cite{bert} and GPT-3 utilize a learnable APE, in which the position embedding $\textbf{P}\!\in\mathbb{R}^{L\times d_{model}}$ for each absolute position is learned along with the model.}

\subsection{Relative Position Embedding}

\textcolor{black}{Instead of encoding absolute position, RPE encodes the distance between each pair of elements in the input sequence. RPE commonly learned a position bias for each relative position, which is then involved in calculating the attention matrix.}

% \textcolor{black}{RPE considers relative positions between frames and there have been several RPE methods explored to inject relative position information into Transformers. RPE often works on raw attention matrix with summation before Softmax normalization.}

\textbf{T5-RPE}~\cite{T5} \textcolor{black}{first bucketize relative distances $i-j$ between the $i$-th and $j$-th elements with a logarithmic bucket function, and the relative positions in the same bucket share the same scalar position bias across layers. A bucket of 32 parameters $\textbf{B}$ is learned for each attention head and the relative position bias $\textbf{P}_{i,j}$ is assigned as}
% \textbf{T5-RPE}~\cite{T5} \textcolor{black}{first exploits a log-binning strategy to split the relative positions $i-j$ between frames at positions $i$ and $j$ into a fixed number of buckets, and the same scalar position bias is shared for the positions within the same bucket across Transformer layers. For each attention head, T5-RPE involves a bucket of 32 learnable parameters $\textbf{B}$, and the position bias is assigned as follows:}
% \begin{equation}
% \textbf{P}_{i, j}\!=\left\{\begin{array}{lr}
% \textbf{B}[\min \left(15,8+\left\lfloor\frac{\log ((|i-j|) / 8)}{\log (128 / 8)} \cdot 8\right\rfloor\right)+16], & i-j \leq-8 \\
% \textbf{B}[|i-j|+16], & -8<i-j<0 \\
% \textbf{B}[i-j], & 0 \leq i-j<8 \\
% \textbf{B}[\min \left(15,8+\left\lfloor\frac{\log ((i-j) / 8)}{\log (128 / 8)} \cdot 8\right\rfloor\right)], & i-j \geq 8
% \end{array}\right.
% \end{equation}
\textcolor{black}{
\begin{equation}
\setlength{\abovedisplayskip}{2pt}
\setlength{\belowdisplayskip}{2pt}
\textbf{P}_{i,j}\!= 
\begin{cases}
\textbf{B}[\min(15,8\!+\!\lfloor\frac{\log ((|i-j|)/8)}{\log(128/8)} \cdot 8\rfloor)\!+\!16],  \,\,\,\,\,\, i\!-j\! \leq -\!8 \\
\textbf{B}[|i-j|+16], \qquad \qquad \qquad \qquad \quad \,\,\,\,\, \!{-8}\!<\! i\!-\!j\!<\!0\\ 
\textbf{B}[i-j], \qquad \qquad \qquad \qquad \qquad \qquad \,\,\,\quad 0\!\leq\!i\!-\!j\!<\!8\\ 
\textbf{B}[\min(15,8\!+\!\lfloor\frac{\log ((i-j) / 8)}{\log (128 / 8)} \cdot 8\rfloor)], \ \ \ \, \,\quad \quad \quad \, \,\, i\!-\!j\!\geq\!8  
\end{cases}
\end{equation}
where $\lfloor\cdot\rfloor$ denotes the flooring operation.}

\textbf{KERPLE}~\cite{chi2022kerple} \textcolor{black}{kernelizes positional differences by using conditionally positive definite kernels. There are two variants of KERPLE, the power variant and the logarithmic variant, with the logarithmic variant showcasing a better performance:
\begin{equation}\label{kerple}
\setlength{\abovedisplayskip}{2pt}
\setlength{\belowdisplayskip}{2pt}
    \textbf{P}_{i,j} = -r_{1}\cdot \text{log}(1+r_{2}|i-j|)
\end{equation}
where $r_{1}, r_{2}\!>\!0$ denote learnable scalars in each head.}

\section{\textcolor{black}{Speech Enhancement with Position-aware Transformer}}\label{sec:4}

\begin{figure}[!bp]
\vspace{-1.9em}
\centering
\includegraphics[width=0.63\columnwidth]{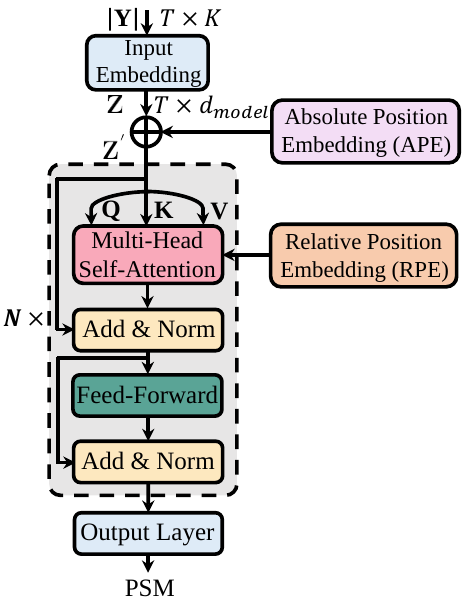}
% \vspace{-0.5em}
\caption{\textcolor{black}{Diagram of position-aware Transformer model for speech enhancement. $\oplus$ represents the element-wise addition.}}
\label{fig2}
\vspace{-1.5em}
\end{figure}

\subsection{Model Architecture}
\textcolor{black}{In Fig.~\ref{fig2}, we depict the position-aware Transformer backbone network. The input to the network is the STFT magnitude spectrum of the noisy mixture, denoted by $|\textbf{Y}|\in\mathbb{R}^{T\times K}$, where $T$ and $K$ respectively represent the number of time frames and frequency bands. The input embedding layer is a fully connected (FC) layer preactivated by frame-wise layer normalization (LN) followed by the ReLU function, which projects the input into a latent embedding, denoted as $\textbf{Z}\in\mathbb{R}^{T\times d_{model}}$. In APE methods, the position embedding $\textbf{P}\in\mathbb{R}^{T\times d_{model}}$ is added to $\textbf{Z}$ to form the input $\textbf{Z}^{\prime}$ to $N$ stacked Transformer layers. Each Transformer layer comprises two sub-layers, an MHSA module and a two-layer feedforward network (FFN). A residual connection is applied around each sub-layer, followed by LN. The output layer is an FC layer with a sigmoidal function, which generates the estimated PSM.}

\subsection{Position-Aware Self-Attention}\label{sec:4.2}

\textcolor{black}{Rather than incorporating position information in the input embedding (APE), RPE integrates relative position information in self-attention. The position-aware MHSA employs $h$ attention heads to facilitate the model to jointly attend to different aspects of information. For each attention head $\textbf{Head}^{i}$ ($i\in\{1,..,h\}$), given a hidden representation $\textbf{U}\in \mathbb{R}^{T\times d_{model}}$ as the input, it is firstly transformed into the queries, keys, and values with three separate project matrices ($\textbf{W}_{Q}^{i}\!\in \mathbb{R}^{d_{\text{model}}\times d_{k}}$, $\textbf{W}_{K}^{i}\in \mathbb{R}^{d_{\text{model}}\times d_{k}}$, $\textbf{W}_{V}^{i}\in\mathbb{R}^{d_{\text{model}}\times d_{v}}$): $\textbf{Q}^{i}\!=\textbf{U}\textbf{W}^{i}_{Q}$, $\textbf{K}^{i}\!=\textbf{U}\textbf{W}^{i}_{K}$, $\textbf{V}^{i}\!=\textbf{U}\textbf{W}^{i}_{V}$, with dimensions of $d_{k}$, $d_{k}$, and $d_{v}$, respectively. The attention scores are computed with the scaled dot-product in each head, where the relative position bias matrix $\textbf{P}^{i}\in\mathbb{R}^{T\times T}$ is integrated into self-attention after query-key dot product:
\begin{equation}
\setlength{\abovedisplayskip}{2pt}
\setlength{\belowdisplayskip}{2pt}
    \textbf{Head}^{i} = \text{Softmax}\left(\frac{\textbf{Q}^{i}\textbf{K}^{i \top}}
    {\sqrt{d_{k}}} + \textbf{P}^{i} \right)\textbf{V}^{i}.
\end{equation}
where $d_{k}\!=\!d_{v}\!=\!d_{model}\!/\!{h}$. Then, the outputs of all $h$ attention heads are concatenated together and passed through a linear feed-forward layer, which is formulated as follows:
\begin{equation}
\setlength{\abovedisplayskip}{5pt}
\setlength{\belowdisplayskip}{5pt}
    \text{MHSA}\left(\textbf{Q}, \textbf{K}, \textbf{V}\right) = \text{Concat}\{\textbf{Head}^{1},...,\textbf{Head}^{h}\}\textbf{W}_{O}
\end{equation}
where $\textbf{W}^{O}\!\in\mathbb{R}^{d_{\text{model}}\times d_{\text{model}}}$ is the linear projection matrix. A two-layer FNN is further applied to the output of the position-ware MHSA sub-layer for two linear transformations.}

\section{Experiments} \label{sec:5}
\subsection{Datasets and Feature Extraction}

\textcolor{black}{The training data consists of $28\,539$ clean speech utterances from the LibriSpeech \textit{train-clean-100} corpus~\cite{panayotov2015librispeech}, with approximately 100 hours. We collect noise recordings from the QUT-NOISE dataset~\cite{dean2010qut}, the Environmental Noise dataset~\cite{saki2016smartphone}, the noise recordings in MUSAN dataset~\cite{snyder2015musan}, the RSG-10 dataset, the UrbanSound dataset~\cite{Urban}, the Nonspeech dataset~\cite{hu2010tandem}, and the colored noise recordings~\cite{deepmmse}. Four noise sources are drawn from collected noise data for evaluation: \textit{F16}, \textit{factory welding}, and \textit{voice babble} from the RSG-10 dataset, and \textit{street music} from the UrbanSound dataset. This results in a noise set comprising $6\,809$ noise clips. For validation experiments, we randomly draw $1\,000$ noise recordings and clean speech utterances to create a validation set of $1\,000$ noisy mixtures, where each clean speech utterance is mixed with a random clip from one noise recording at an SNR randomly sampled from -10 to 20 dB (in 1 dB increments). All audio signals are with a sampling frequency of 16 kHz. We extract the STFT magnitude spectrum using a square-root-Hann window of length 32 ms (512 time samples) with a frame shift of 16 ms (256 time samples). }

\textbf{Length Generalization}. \textcolor{black}{We segment the clean speech utterances in the training set into 1-second (1s) and 2-second (2s) speech clips for training, respectively. For evaluation experiments, we generate six test sets, each comprising 400 noisy mixtures of 1s, 2s, 5s, 10s, 15s, and 20s, respectively. The process of generating mixtures is as follows: For each of the four test noise recordings, we randomly take 20 clean speech (over durations of 1s, 2s, 5s, 10s, 15s, and 20s, respectively) from the LibriSpeech \textit{test-clean-100} corpus. A random clean clip (1s, 2s, 5s, 10s, 15s, and 20s in length) is mixed with a random noise clip (with the same length) from the noise recording, across SNRs from -5 and 15 dB, in 5 dB increments. The validation lengths have a broad range (diversity) and most of the speech utterances are more than four times (4s or 8s) the length of the training set (1s or 2s).
}

\subsection{Implementation Details}

\textcolor{black}{To explore the impact of position embeddings in length generalization, we employ a Transformer backbone without position embedding (No-Pos) as a base. The Transformer backbone contains $N\!=\!4$ Transformer layers, with the parameter configurations as in~\cite{ripple}: $h\!=\!8$, $d_{model}\!=\!256$, and $d_{f\!f}\!=\!1024$. Our experiments explore length generalization across four commonly used position embeddings, Sniusoidal~\cite{transformer}, learnable-APE~\cite{kenton2019bert}, T5-RPE~\cite{T5}, and KERPLE~\cite{chi2022kerple}. The original KERPLE is proposed for causal language modeling. In this work, a non-causal KERPLE is derived to investigate length generalization.}

\textbf{Training Strategy}. \textcolor{black}{For each training iteration, we select 10 clean speech utterances from the training data and segment them into 1s and 2s clean speech clips, respectively. The speech clips are dynamically mixed with noise clips during the training stage. Specifically, each clean clip in a mini-batch is mixed with a random clip from a random noise recording at a random SNR sampled between -10 and 20 dB (in 1 dB increments). We employ mean-square error (MSE) as the objective function for mask estimation. The Adam algorithm is adopted for the optimization of gradients, with the hyper-parameter settings as~\cite{transformer}, $\beta_{1}\!=\!0.9$, $\beta_{2}\!=\!0.98$, and $\epsilon\!=\!1\times10^{-9}$. The models are trained for 150 epochs. The gradient clipping technique is utilized to keep the gradients between -1 and 1. A warm-up scheduler~\cite{mhanet,transformer} is employed to dynamically adapt the learning rate, which is formulated as: $lr = d_{model}^{-0.5}\cdot \textrm{min} \left(n\_itr^{-0.5}, n\_itr \cdot w\_itr^{-1.5}\right)$, where $n\_itr$ and $w\_itr$ denote the number of training iterations and warm-up stage iterations, respectively. Following the study~\cite{tfaj}, $w\_itr$ is set to $40\,000$. We employ an NVIDIA Tesla P100-PCIe-16GB graphics processing unit to run our experiments.}

\begin{figure}[!htbp]
% \vspace{-0.8em}
\centering
\begin{subfigure}[!htbp]{0.49\columnwidth}\centerline{\includegraphics[width=\columnwidth]{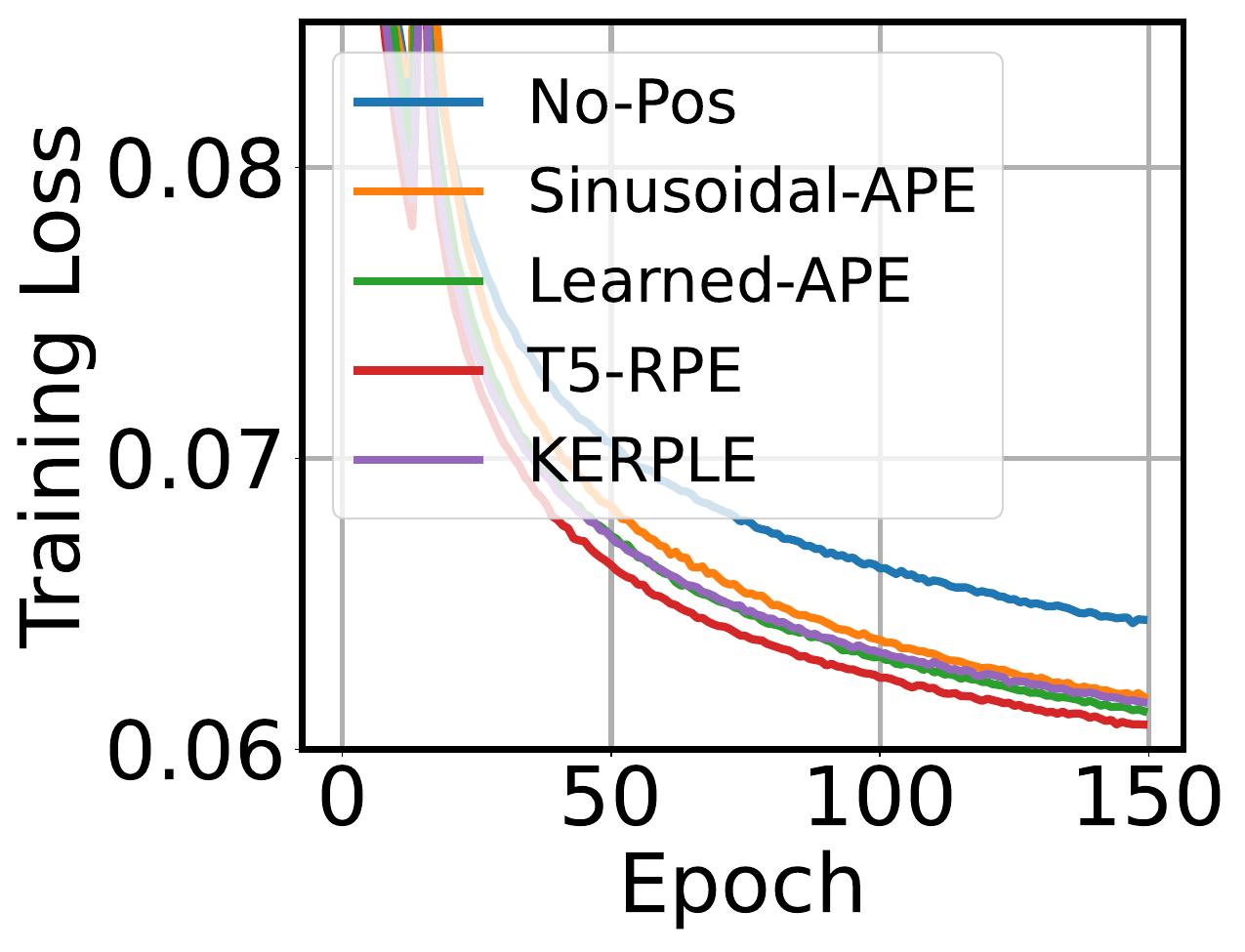}}
\caption{}
\label{fig3:1}
\end{subfigure}
% \hspace{10mm}
\begin{subfigure}[!htbp]{0.49\columnwidth}
\centerline{\includegraphics[width=\columnwidth]{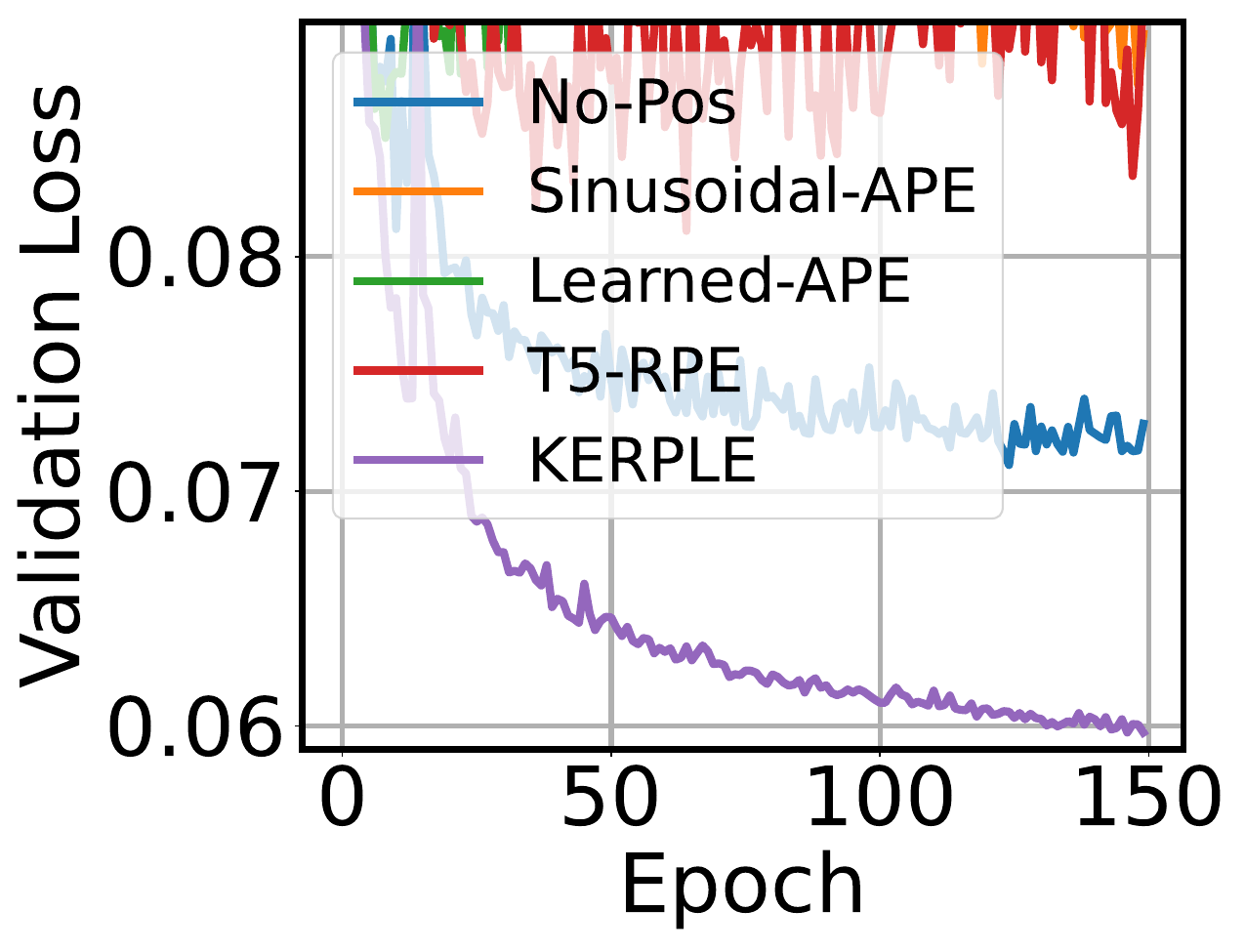}}
\caption{}
\label{fig3:2}
\end{subfigure}
% \vspace{-0.8em}
\caption{The (a) training loss and (b) validation loss of the models trained using speech of length 1s.}
\label{fig3}
\vspace{-2.0em}
\end{figure}

\subsection{Training and Validation Loss}

\begin{figure}[!tbp]
% \vspace{-0.8em}
\centering
\begin{subfigure}[!htbp]{0.49\columnwidth}\centerline{\includegraphics[width=\columnwidth]{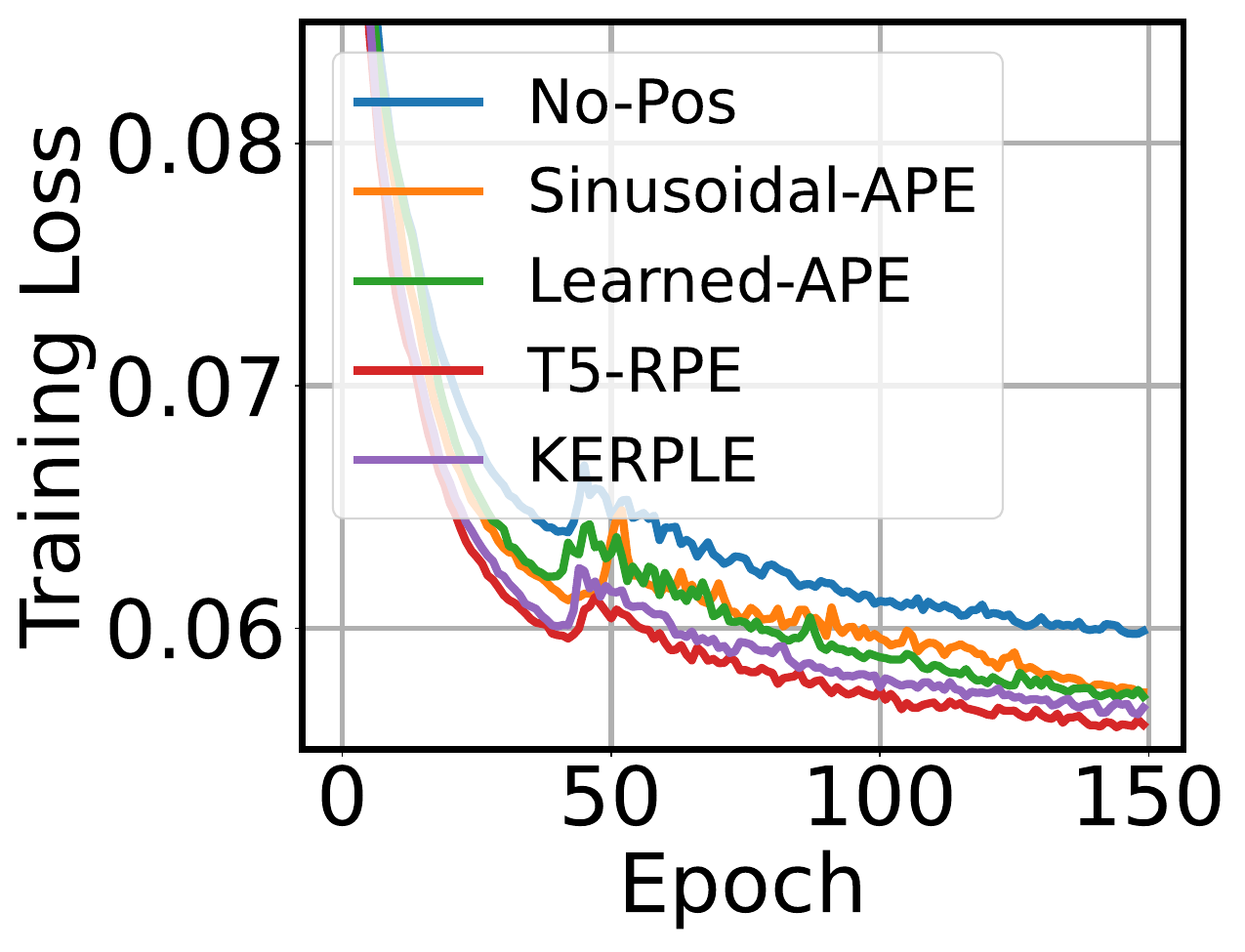}}
\caption{}
\label{fig4:1}
\end{subfigure}
% \hspace{10mm}
\begin{subfigure}[!htbp]{0.49\columnwidth}
\centerline{\includegraphics[width=\columnwidth]{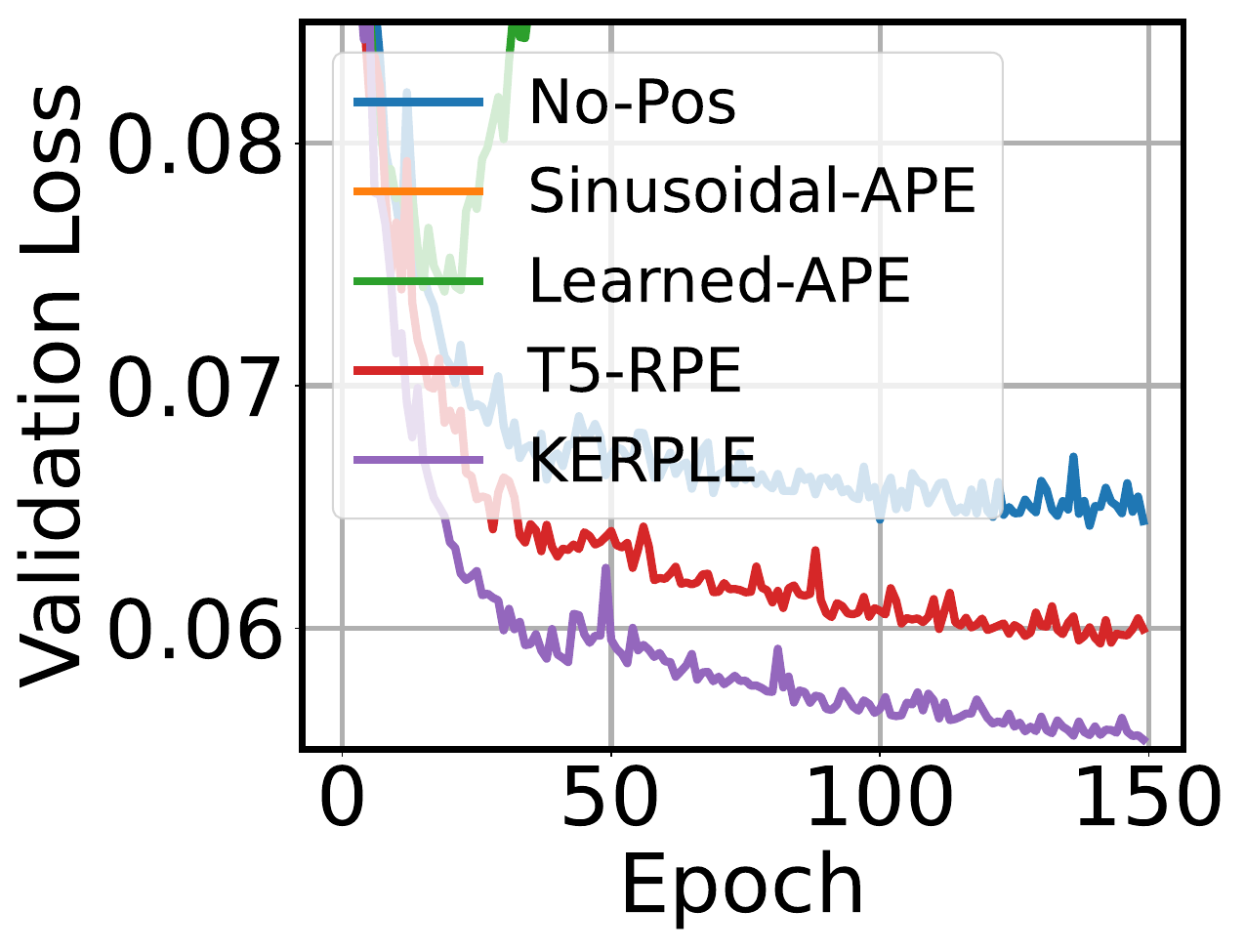}}
\caption{}
\label{fig4:2}
\end{subfigure}
\vspace{-0.8em}
\caption{The (a) training loss and (b) validation loss of the models trained using speech of length 2s.}
\label{fig4}
\vspace{-1.0em}
\end{figure}

\textcolor{black}{Fig.~\ref{fig3}-\ref{fig4} give the train and validation loss curves of the models trained on 1s and 2s speech clips, respectively. It can be found that all the models converge well on the training data. Sinusoidal-APE and Learned-APE fail to perform length generalization. In comparison to No-Pos, they consistently achieve lower training loss but much higher validation loss. We can also observe significant length generalization issues with T5-RPE trained using 1s speech utterances. In contrast, KERPLE consistently demonstrates remarkable length generalization capability, with substantially lower training loss and validation loss compared to No-Pos. Overall, among these methods, RPE methods show a better length generalization property than APE.}

\input{Train1S}

\input{Train2S}
\subsection{Results and Analysis}

\textcolor{black}{We evaluate the performance of models with five objective metrics, i.e., perceptual evaluation of speech quality (PESQ)~\cite{pesq}, extended short-time objective intelligibility (ESTOI)~\cite{jensen2016algorithm}, and three composite metrics (CSIG, CBAK, and COVL)~\cite{hu2007evaluation}.}

\textcolor{black}{Tables~\ref{tab1}-\ref{tab2} respectively report the evaluation results of models trained on 1s and 2s speech clips in terms of PESQ, ESTOI, CSIG, CBAK, and COVL, across five test lengths. For matched test lengths (1s or 2s), all the position embedding methods provide substantial improvements over the No-Pos and exhibit similar performance, with T5-RPE showcasing the best results, which is consistent with the loss trends in Fig.~\ref{fig3}-\ref{fig4}. For the test lengths from 5s to 20s, among these methods, overall KERPLE exhibits the best length generalization property. In the 10s and 20s test lengths case (1s training length, shown in Table~\ref{tab1}), KERPLE attain the gains of 0.24 and 0.25, 4.73\% and 5.38\%, 0.23 and 0.25, 0.18 and 0.20, 0.23 and 0.25, on PESQ, ESTOI, CSIG, CBAK, and COVL, respectively, over No-Pos. Sinusoidal-APE and Learned-APE exhibit similar length generalization issues, with substantially lower performance scores compared to No-Pos. Taking the 10s test length case (2s training length), compared to No-Pos, Sinusoidal-APE and Learned-APE show decreases of 0.58 and 0.53, and 12.12\% and 10.67\% in PESQ and ESTOI, respectively. T5-RPE also shows a similar length generalization issue for 1s training length. }

\vspace{-0.5em}
\section{Conclusion}\label{sec:6}

\textcolor{black}{In this paper, we establish and explore the length generalization problem in speech enhancement models with Transformers. We introduce four position embedding methods to enable Transformer-based speech enhancement models with the property to generalize to longer utterances. The comprehensive experiment results in five metrics underscore the superiority of relative position embedding (RPE) over absolute position embedding (APE) in terms of length generalization. APE methods, i.e., Sinusoidal-APE and Learned-APE are not able to generalize from shorter speech to long ones. We believe our work sets the stage for future research on length generalization for speech enhancement and some other speech processing tasks with Transformers.}

\bibliographystyle{IEEEtran}
\bibliography{mybib}

\end{document}

%% file: Train1S.tex
\begin{table}[!t]
    \centering
    \scriptsize
    \small
    \def\arraystretch{0.91}
    \setlength{\tabcolsep}{1.3pt}
    \setlength{\abovetopsep}{0pt}
    \setlength\belowbottomsep{0pt} 
    \setlength\aboverulesep{0pt} 
    \setlength\belowrulesep{0pt}
    % \vspace*{-0.1in}
    % \footnotesize	
    % \caption{Average CSIG scores for each SNR level and the highest CSIG scores are highlighted in boldface.}
    \caption{\textcolor{black}{The comparison results of PESQ, ESTOI (in \%), CSIG, CBAK, and COVL. All the models are trained using 1s noisy-clean speech pairs and tested on noisy mixtures of lengths 1s, 5s, 10s, 15s, and 20s.}}
    \label{tab1}
    \vspace*{-0.1in}
    \begin{tabular}{@{}cl||ccccc@{}}
        \toprule[1.5pt]
         % &  & \multicolumn{5}{c}{\textbf{Input SNR (dB)}}\\  
         &  & \multicolumn{5}{c}{\textbf{Metrics}}\\  
        % \cline{3-7}
        Test Len. & Model & PESQ & ESTOI & CSIG & CBAK & COVL\\  
        % \hline
        % Objective & Network & -5 & 0 & 5 & 10 & 15 \\
        % \hline
        \hline
        % \hline
        \multirow{6}{*}{1s}
        & Noisy    & 1.88 & 54.26 & 2.33 & 1.86 & 1.71 \\
        % \cline{2-7}
        % \cdashline{2-7}
        & No-Pos           & 2.59 & 68.17 & 3.18 & 2.60 & 2.48  \\
        & Sinusoidal-APE   & 2.70 & 70.96 & 3.29 & 2.68 & 2.59  \\
        & Learned-APE      & 2.72 & 71.12 & 3.30 & 2.68 & 2.60  \\
        % & Gauss-Bias   & 2.60 & 68.19 & 3.20 & 2.61 & 2.49  \\
        & T5-RPE           & \textbf{2.73} & \textbf{71.60} & \textbf{3.34} & \textbf{2.70} & \textbf{2.63}  \\
        % & TISA             & \textbf{2.73} & \textbf{71.67} & 3.33 & \textbf{2.70} & 2.62  \\
        % & \textcolor{black}{DA-Bias}   & 2.71 & 70.91 & 3.29 & 2.68 & 2.58  \\
        & \textcolor{black}{KERPLE}       & 2.71 & 70.51 & 3.29 & 2.69 & 2.60  \\
        % & LearnLin     & 2.72 & 70.97 & 3.31 & 2.68 & 2.61 \\
        \hline
        % \hline
        \multirow{6}{*}{5s}
        & Noisy    & 1.88 & 54.26 & 2.33 & 1.86 & 1.71 \\
        % \cline{2-7}
        % \cdashline{2-7}
        & No-Pos       & 2.49 & 66.02 & 3.10 & 2.50 & 2.38  \\
        & Sinusoidal-APE   & 2.42 & 62.50 & 2.94 & 2.41 & 2.26  \\
        % & BERT-Pos     & 2.44 & 63.48 & 3.07 & 2.45 & 2.34  \\
        & Learned-APE     & 2.11 & 55.49 & 2.61 & 2.21 & 1.96  \\
        % & Gauss-Bias   & 2.55 & 67.26 & 3.18 & 2.55 & 2.46  \\
        & T5-RPE      & 2.52 & 65.61 & 3.12 & 2.48 & 2.40  \\
        % & TISA         & 2.71 & 70.57 & \textbf{3.36} & 2.66 & \textbf{2.63}  \\
        % & \textcolor{black}{DA-Bias}   & 2.70 & 70.11 & 3.34 & 2.67 & 2.61  \\
        & \textcolor{black}{KERPLE}   & \bf 2.69 & \bf 70.27 & \bf 3.32 & \bf 2.64 & \bf 2.59  \\
        % & LearnLin     & \textbf{2.72} & \textbf{71.03} & 3.34 & \textbf{2.67} & 2.62 \\
        \hline
        % \hline
        
        \multirow{6}{*}{10s}
        & Noisy        & 1.92 & 53.19 & 2.40 & 1.87 & 1.75 \\
        % \cline{2-7}
        % \cdashline{2-7}
        & No-Pos       & 2.46 & 63.89 & 3.09 & 2.45 & 2.36  \\
        & Sinusoidal-APE   & 2.38 & 60.17 & 2.92 & 2.35 & 2.22  \\
        % & BERT-Pos     & 2.35 & 59.44 & 2.97 & 2.36 & 2.25  \\
        & Learned-APE     & 2.08 & 53.44 & 2.54 & 2.15 & 1.91  \\
        % & Gauss-Bias   & 2.57 & 65.66 & 3.21 & 2.55 & 2.47  \\
        & T5-RPE      & 2.42 & 61.01 & 3.01 & 2.37 & 2.29  \\
        % & TISA         & 2.67 & 67.65 & 3.30 & 2.60 & 2.58  \\
        % & \textcolor{black}{DA-Bias}   & 2.67 & 67.45 & 3.30 & 2.62 & 2.58  \\
        & \textcolor{black}{KERPLE}   & \bf 2.70 & \bf 68.62 & \bf 3.32 & \bf 2.63 & \bf 2.59  \\
        % & LearnLin     & \textbf{2.73}& \textbf{69.49} & \textbf{3.37}& \textbf{2.67}& \textbf{2.64} \\
        \hline
        % \hline
        
        \multirow{6}{*}{15s}
        & Noisy        & 1.90 & 52.68 & 2.32 & 1.83 & 1.71 \\
        % \cdashline{2-7}
        & No-Pos       & 2.43 & 63.18 & 3.01 & 2.39 & 2.30  \\
        & Sinusoidal-APE   & 2.35 & 58.24 & 2.83 & 2.28 & 2.15  \\
        % & BERT-Pos     & 2.30 & 57.70 & 2.88 & 2.28 & 2.17  \\
        & Learned-APE     & 2.05 & 51.42 & 2.46 & 2.09 & 1.85  \\
        % & Gauss-Bias   & 2.53 & 65.24 & 3.14 & 2.49 & 2.41  \\
        & T5-RPE      & 2.35 & 58.89 & 2.89 & 2.29 & 2.19  \\
        % & TISA         & 2.58 & 65.98 & 3.18 & 2.51 & 2.47  \\
        % & \textcolor{black}{DA-Bias}   & 2.59 & 66.18 & 3.21 & 2.54 & 2.49  \\
        & \textcolor{black}{KERPLE}   & \bf 2.67 & \bf 68.24 & \bf 3.25 & \bf 2.58 & \bf 2.54  \\
        % & LearnLin     & \textbf{2.71} & \textbf{69.1}5 & \textbf{3.30} & \textbf{2.63} & \textbf{2.59}  \\
        \hline
        % \hline
        
        \multirow{6}{*}{20s}
        & Noisy        & 1.91 & 51.42 & 2.36 & 1.82 & 1.72    \\
        % \cline{2-7}
        % \cdashline{2-7}
        & No-Pos       & 2.43 & 61.50 & 3.05 & 2.41 & 2.32  \\
        & Sinusoidal-APE   & 2.39 & 56.71 & 2.87 & 2.29 & 2.19  \\
        % & BERT-Pos     & 2.30 & 55.92 & 2.90 & 2.27 & 2.18  \\
        & Learned-APE     & 2.03 & 50.19 & 2.44 & 2.08 & 1.83  \\
        % & Gauss-Bias   & 2.55 & 64.11 & 3.19 & 2.52 & 2.45  \\
        & T5-RPE      & 2.40 & 57.02 & 2.88 & 2.28 & 2.18  \\
        % & TISA         & 2.60 & 64.07 & 3.20 & 2.51 & 2.48  \\
        % & \textcolor{black}{DA-Bias}   & 2.59 & 63.87 & 3.21 & 2.53 & 2.49  \\
        & \textcolor{black}{KERPLE}   & \bf{2.68 }& \bf{66.88} & \bf{3.30 }& \bf{2.61 }& \bf{2.57 } \\
        % & LearnLin     & \textbf{2.73} & \textbf{67.93} & \textbf{3.36} & \textbf{2.65} & \textbf{2.63}  \\
        % \hline
        \toprule[1.5pt]
    \end{tabular}
    \vspace{-1.9em}
\end{table}

%% file: Train2S.tex
\begin{table}[!t]
    \centering
    \scriptsize
    \small
    \def\arraystretch{0.92}
    \setlength{\tabcolsep}{1.2pt}
    \setlength{\abovetopsep}{0pt}
    \setlength\belowbottomsep{0pt} 
    \setlength\aboverulesep{0pt} 
    \setlength\belowrulesep{0pt}
    % \footnotesize	
    % \caption{Average CSIG scores for each SNR level and the highest CSIG scores are highlighted in boldface.}
    \caption{\textcolor{black}{The comparison results of PESQ, ESTOI (in \%), CSIG, CBAK, and COVL. All the models are trained using 2s noisy-clean speech pairs and tested on noisy mixtures of lengths 2s, 5s, 10s, 15s, and 20s.}}
    \label{tab2}
    \vspace*{-0.1in}
    \begin{tabular}{@{}cl||ccccc@{}}
        % \hline
        \toprule[1.5pt]
         % &  & \multicolumn{5}{c}{\textbf{Input SNR (dB)}}\\  
         &  & \multicolumn{5}{c}{\textbf{Metrics}}\\  
        % \cline{3-7}
        Test Len. & Model & PESQ & ESTOI & CSIG & CBAK & COVL\\  
        % \hline
        % Objective & Network & -5 & 0 & 5 & 10 & 15 \\
        % \hline
        \hline
        % \hline
        \multirow{6}{*}{2s}
        & Noisy    & 1.88 & 53.94 & 2.36 & 1.87 & 1.74 \\
        % \cline{2-7}
        % \cdashline{2-7}
        & No-Pos       & 2.58 & 67.19 & 3.23 & 2.61 & 2.50  \\
        & Sinusoidal-APE   & 2.71 & 70.23 & 3.33 & 2.69 & 2.61  \\
        & Learned-APE     & 2.70 & 70.55 & 3.33 & 2.70 & 2.62  \\
        % & Gauss-Bias   & 2.60 & 67.49 & 3.24 & 2.62 & 2.52  \\
        & T5-RPE      & \textbf{2.74} & \bf{71.15} & \bf{3.37 }& \bf{2.71 }& \textbf{2.65}  \\
        % & TISA         & \textbf{2.74} & \textbf{71.33} & \textbf{3.38} & 2.71 & \textbf{2.65}  \\
        % & \textcolor{black}{DA-Bias}   & 2.73 & 70.94 & 3.37 & 2.71 & 2.65  \\
        & \textcolor{black}{KERPLE}    & 2.71 & 70.34 & 3.34 & 2.69 & 2.62  \\
        % & LearnLin     & 2.72 & 70.81 & 3.36 & \textbf{2.72} & 2.64  \\
        \hline
        % \hline
        \multirow{6}{*}{5s}
        & Noisy    & 1.88 & 54.26 & 2.33 & 1.86 & 1.71 \\
        % \cline{2-7}
        % \cdashline{2-7}
        & No-Pos       & 2.56 & 67.40 & 3.19 & 2.57 & 2.47  \\
        & Sinusoidal-APE   & 1.89 & 53.44 & 2.31 & 1.91 & 1.71  \\
        % & BERT-Pos     & 2.60 & 68.26 & 3.24 & 2.59 & 2.52  \\
        & Learned-APE     & 2.06 & 59.48 & 2.61 & 2.10 & 1.92  \\
        % & Gauss-Bias   & 2.61 & 68.40 & 3.25 & 2.60 & 2.52  \\
        & T5-RPE      & \bf 2.74 & 71.29 & \bf 3.39 & \bf 2.70 & \bf 2.66  \\
        % & TISA         & \textbf{2.76} & 71.80 & \textbf{3.41} & 2.69 & \textbf{2.68}  \\
        % & \textcolor{black}{DA-Bias}   & 2.74 & 71.86 & \textbf{3.41} & 2.71 & 2.67  \\
        & \textcolor{black}{KERPLE}    & \bf 2.74 & \bf 71.38 & 3.37 & 2.67 & 2.64  \\
        % & LearnLin     & 2.75 & \textbf{71.88} & 3.40 & \textbf{2.71} & 2.67 \\
        \hline
        % \hline
        
        \multirow{6}{*}{10s}
        & Noisy    & 1.92 & 53.19 & 2.40 & 1.87 & 1.75 \\
        % \cline{2-7}
        % \cdashline{2-7}
        & No-Pos       & 2.52 & 65.35 & 3.19 & 2.53 & 2.45  \\
        & Sinusoidal-APE   & 1.94 & 53.23 & 2.37 & 1.92 & 1.76  \\
        % & BERT-Pos     & 2.49 & 63.39 & 3.14 & 2.48 & 2.39  \\
        & Learned-APE     & 1.99 & 54.68 & 2.47 & 1.97 & 1.82  \\
        % & Gauss-Bias   & 2.61 & 66.90 & 3.27 & 2.61 & 2.54  \\
        & T5-RPE      & 2.69 & 67.99 & 3.35 & 2.64 & 2.62  \\
        % & TISA         & 2.72 & 69.28 & 3.38 & 2.65 & 2.65  \\
        % & \textcolor{black}{DA-Bias}   & 2.73 & 69.54 & 3.41 & 2.68 & 2.67  \\
        & \textcolor{black}{KERPLE}    & \bf 2.74 & \bf 69.95 & \bf 3.39 & \bf 2.67 & \bf 2.65  \\
        % & LearnLin     & \textbf{2.75} & \textbf{70.40} & \textbf{3.42} & \textbf{2.72} & \textbf{2.69}  \\
        \hline
        % \hline
        
        \multirow{6}{*}{15s}
        & Noisy        & 1.90 & 52.68 & 2.32 & 1.83 & 1.71 \\
        % \cdashline{2-7}
        & No-Pos       & 2.47 & 64.44 & 3.10 & 2.46 & 2.37  \\
        & Sinusoidal-APE   & 1.92 & 52.78 & 2.28 & 1.88 & 1.70  \\
        % & BERT-Pos     & 2.42 & 60.92 & 3.02 & 2.38 & 2.29  \\
        & Learned-APE     & 1.96 & 52.49 & 2.40 & 1.89 & 1.76  \\
        % & Gauss-Bias   & 2.59 & 66.52 & 3.22 & 2.56 & 2.49  \\
        & T5-RPE      & 2.60 & 66.62 & 3.24 & 2.54 & 2.51  \\
        % & TISA         & 2.65 & 68.18 & 3.27 & 2.56 & 2.55  \\
        % & \textcolor{black}{DA-Bias}   & 2.69 & 68.85 & 3.36 & 2.64 & 2.63  \\
        & \textcolor{black}{KERPLE}    & \bf 2.72 & \bf 69.59 & \bf 3.35 & \bf 2.64 & \bf 2.62  \\
        % & LearnLin     & \textbf{2.74} & \textbf{70.18} & \textbf{3.38} & \textbf{2.68} & \textbf{2.65}  \\
        \hline
        % \hline
        
        \multirow{5}{*}{20s}
        & Noisy        & 1.91 & 51.42 & 2.36 & 1.82 & 1.72    \\
        % \cline{2-7}
        % \cdashline{2-7}
        & No-Pos       & 2.49 & 62.73 & 3.13 & 2.47 & 2.40  \\
        & Sinusoidal-APE   & 1.91 & 51.37 & 2.33 & 1.86 & 1.72  \\
        % & BERT-Pos     & 2.43 & 58.79 & 3.04 & 2.38 & 2.30  \\
        & Learned-APE     & 1.94 & 51.55 & 2.40 & 1.87 & 1.75  \\
        % & Gauss-Bias   & 2.62 & 65.50 & 3.27 & 2.59 & 2.53  \\
        & T5-RPE      & 2.61 & 64.20 & 3.26 & 2.54 & 2.52  \\
        % & TISA         & 2.66 & 66.12 & 3.31 & 2.58 & 2.58  \\
        % & \textcolor{black}{DA-Bias}   & 2.71 & 67.27 & 3.39 & 2.64 & 2.64  \\
        & \textcolor{black}{KERPLE}    & \bf 2.74 & \bf 68.36 & \bf 3.39 & \bf 2.65 & \bf 2.65  \\
        % & LearnLin     & \textbf{2.78} & \textbf{68.85} & \textbf{3.43} & \textbf{2.70} & \textbf{2.69}  \\
        % \hline
        \toprule[1.5pt]
    \end{tabular}
    \vspace{-1.9em}
\end{table}